# Angular Momentum Exchange Between Light and Material Media Deduced from the Doppler Shift

Masud Mansuripur

College of Optical Sciences, The University of Arizona, Tucson, Arizona 85721

**Abstract**. Electromagnetic waves carry energy as well as linear and angular momenta. When a light pulse is reflected from, transmitted through, or absorbed by a material medium, energy and momentum (both linear and angular) are generally exchanged, while the total amount of each entity remains intact. The extent of such exchanges between light and matter can be deduced, among other methods, with the aid of the Doppler shift phenomenon. The main focus of the present paper is on the transfer of angular momentum from a monochromatic light pulse to spinning objects such as a mirror, an absorptive dielectric, or a birefringent plate. The fact that individual photons of frequency $\omega_o$ carry energy in the amount of $\hbar\omega_o$, where $\hbar$ is Planck's reduced constant, enables one to relate the Doppler shift to the amount of energy exchanged. Under certain circumstances, the knowledge of exchanged energy leads directly to a determination of the momentum transferred from the photon to the material body, or vice versa.

**1. Introduction**. Interactions of light with material media involve exchanges of energy, momentum, and angular momentum (AM). The generalized Lorentz law of force, also known as the Einstein-Laub force-density equation, together with the corresponding equations for torque and electro-magnetic (EM) momentum densities, can be used to analyze the transfer of momentum from radiation to matter, or vice versa [1,2]. A powerful alternative method of analyzing such problems is based on the phenomenon of Doppler shift, which relates changes in photon frequencies to the amount of energy and momentum exchanged between light and matter [3-10]. In previous publications, we used the Doppler-shift method to derive expressions for the radiation pressure and the linear momentum of light when submerged mirrors or absorptive media are exposed to EM radiation [11,12]. The goal of the present paper is to extend the technique to an analysis of angular momentum exchange between a monochromatic, circularly-polarized light pulse and various spinning objects.

A homogeneous, isotropic, linear medium is typically specified in terms of its relative permittivity $\varepsilon(\omega)$ and permeability $\mu(\omega)$, where $\omega$ is the excitation frequency. The complex refractive index of the material is then given by $n+\mathrm{i}\kappa = \sqrt{\mu(\omega)\varepsilon(\omega)}$. For a transparent medium, $\kappa \approx 0$ and the real-valued refractive index $n$ is sometimes denoted by $n_p$ and referred to as the phase index. If both $\mu(\omega)$ and $\varepsilon(\omega)$ happen to be real-valued and negative in some range of frequencies, the corresponding refractive index $n_p$ will also be negative; the material is then referred to as a negative-index medium (NIM). A narrow-band light pulse of central frequency $\omega_o$, propagating in a transparent, dispersive medium (be it positive- or negative-index), generally travels at the group velocity $V_g = c/n_g$, where $c$ is the speed of light in vacuum, and $n_g$, the group refractive index, is given by

$$n_g = \mathrm{d}[\omega n_p(\omega)]/\mathrm{d}\omega \big|_{\omega=\omega_o}. \tag{1}$$

Two other parameters often used to describe the EM properties of material media are the impedance $Z(\omega) = \sqrt{\mu/\varepsilon}$ and the admittance $\eta(\omega) = \sqrt{\varepsilon/\mu}$. In transparent materials, both $Z$ and $\eta$ are real-valued and positive, even when the phase refractive index $n_p$ is negative.

Inside a transparent medium, a circularly-polarized photon of frequency $\omega_o$ and energy $\hbar\omega_o$, which, in vacuum, would have linear momentum $\hbar\omega_o/c$ and angular momentum $\pm\hbar$, carries both electromagnetic and mechanical momenta. The EM linear momentum of the photon inside the medium is its Abraham momentum, $p_{EM} = \hbar\omega_o/(n_g c)$, while its total linear momentum is given by

$$p_{total} = p_{EM} + p_{mech} = \tfrac{1}{2}[\sqrt{\mu(\omega_o)/\varepsilon(\omega_o)} + \sqrt{\varepsilon(\omega_o)/\mu(\omega_o)}](\hbar\omega_o/c) = \tfrac{1}{2}[Z(\omega_o) + \eta(\omega_o)](\hbar\omega_o/c). \quad (2)$$

Note that, inside a NIM, a photon's $p_{EM}$ and $p_{total}$ are both positive (i.e., in the direction of propagation), despite the fact that the phase of the corresponding EM wave propagates in the opposite direction [1,12,13]. As for the AM of a circularly-polarized photon inside a transparent medium, the total spin AM remains the same as that in free space, namely, $S_{total} = \pm\hbar$. However, this angular momentum now has an EM component, given by $S_{EM} = \pm\hbar/(n_p n_g)$, and a mechanical component $S_{mech} = S_{total} - S_{EM}$ [1]. In negative-index media, $S_{EM}$ and $S_{total}$ have opposite signs.

In the remainder of this paper, we examine the behavior of spinning mirrors, absorbers, and wave-plates in the presence of circularly-polarized light. A narrow-band light pulse of frequency $\omega_o$ propagating in free space typically contains a large number $N$ of photons, each carrying energy $\hbar\omega_o$, linear momentum $\hbar\omega_o/c$, and spin angular momentum $\pm\hbar$. Our assumption throughout the paper will be that $N$ is large enough to render the behavior of the light pulse classical, that is, in compliance with the usual rules of classical electrodynamics. The stated properties of individual photons (e.g., energy, momentum) should then be taken as the corresponding properties of the light pulse normalized by $N$.

In the next section, we briefly introduce the Doppler-based method of analysis for the linear momentum of light in conjunction with moving mirrors and absorbers. We then proceed to extend this method to an analysis of angular momentum in the presence of spinning mirrors, absorbers, transparent dielectric slabs (isotropic), and transparent wave-plates (birefringent). A rigorous analysis of such problems would require the solution of Maxwell's equations in a rotating coordinate system – an undertaking that goes beyond the scope of the present paper. Our results should, therefore, be recognized for their approximate, non-relativistic nature, and trusted only at small angular velocities.

**2. Linear momentum exchanged upon reflection or absorption of a photon.** Consider a perfect reflector of mass $M_o$ moving at constant velocity $V \ll c$ along the $x$-axis, as shown in Fig. 1, where the thin-film coating is assumed to be a high-conductivity metallic layer. The reflection of a single photon of frequency $\omega_o$ and energy $\hbar\omega_o$ from the mirror will cause the mirror velocity to increase by $\Delta V$, giving it an added momentum $\Delta\boldsymbol{p}_{mirror} = M_o \Delta V \hat{\boldsymbol{x}}$ and an increased kinetic energy $\Delta\mathcal{E}_{kinetic}$, where

$$\Delta\mathcal{E}_{kinetic} = \tfrac{1}{2}M_o(V+\Delta V)^2 - \tfrac{1}{2}M_o V^2 \approx M_o V \Delta V. \quad (3)$$

Considering that the Doppler-shifted photon frequency after reflection is $\omega = \omega_o(1 - 2V/c)$, the kinetic energy gain of the mirror must be equated with $\hbar\Delta\omega = 2\hbar\omega_o V/c$. This yields $M_o \Delta V \approx 2\hbar\omega_o/c$, corresponding to a photon linear momentum (in vacuum) of $\boldsymbol{p}_{photon} = (\hbar\omega_o/c)\hat{\boldsymbol{x}}$, which we know to be correct. If the initial mirror velocity happens to be zero, one can use a similar argument and arrive at the same conclusion, with the caveat that the Doppler shift is now proportional to the *average* mirror velocity, $\tfrac{1}{2}V$, where $V$ is the final velocity of the mirror (i.e., after the photon has been reflected).

Next, consider a perfect absorber, again as shown in Fig. 1, this time with the thin-film layer acting as an anti-reflection coating. (It is sufficient, in fact, to assume that the weakly-absorptive medium is somehow impedance-matched to the incidence medium – free space in the present case – so that no light is reflected at the entrance facet.) For an observer at rest within the absorptive medium, the Doppler-shifted photon entering the medium has frequency $\omega = \omega_o(1 - V/c)$. The reduction in the photon's energy upon entering the absorber is thus $\hbar \Delta \omega = \hbar \omega_o V/c$, with the remainder of its energy being converted to heat without contributing to the kinetic energy of the absorber. Consequently, $\Delta \mathcal{E}_{\text{kinetic}} \approx M_o V \Delta V = \hbar \omega_o V/c$. The photon momentum, which has been transferred to the absorber, is thus found to be $p_{\text{photon}} = M_o \Delta V = \hbar \omega_o/c$, in agreement with the previous result and also with the known photon momentum in vacuum. Once again, one could make the argument for a stationary absorber and arrive at the same conclusion, provided that the Doppler shift of the photon entering the absorber is associated with the *average* velocity ½V, where V is the final velocity of the absorber (i.e., its velocity after the photon has been fully extinguished).

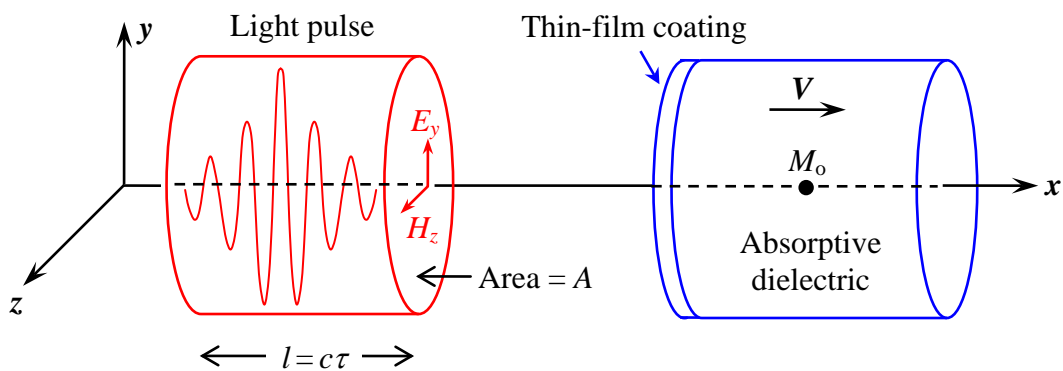

**Fig. 1**. A collimated light pulse of duration $\tau$, cross-sectional area $A$, and center frequency $\omega_o$ propagates along the *x*-axis. The pulse is long enough to qualify as monochromatic, has a sufficiently large cross-sectional area to render it non-diffracting (or collimated), and contains a large number of photons to be considered a classical EM wave-packet. The moving solid object of mass $M_o$ and velocity $V\hat{x}$ with which the light interacts is either an ideal mirror or a perfect absorber, depending on the nature of the thin-film coating at its front facet. If the coating is a high-conductivity metallic layer, the object acts as an ideal mirror. Alternatively, the thin film layer could be an anti-reflection coating that allows the light pulse to enter the substrate and proceed to full extinction within this weakly-absorptive medium. In either case, the object's velocity $V$ must change by some small amount $\Delta V$ in order to preserve the total momentum of the system.

The type of argument employed in the present section may be used to determine the photon momentum, as in the above discussion, or it could be turned around and used to estimate the Doppler shift if the photon momentum happens to be known in advance. In the following sections, we will use the Doppler-based argument in one form or the other, in conjunction with conservation of energy and angular momentum, in order to examine a few situations involving circularly-polarized photons reflecting from, transmitting through, or getting absorbed within various material media.

**3. Circularly-polarized light reflected from spinning flat mirror**. A circularly-polarized photon of frequency $\omega_o$ is reflected from a perfect flat mirror at normal incidence, as shown in Fig. 2. The medium of incidence is free space, and the mirror, whose mass is $M_o$, and whose moment of inertia around the *z*-axis is $I_o$, is spinning around *z* at a constant angular velocity $\Omega$. We assume the mirror is sufficiently massive, and proceed to ignore its kinetic energy $½M_o V^2$ acquired in the wake of the transfer of the linear momentum $2\hbar \omega_o/c$ from the incident photon. One expects the polarization state of the photon to switch, upon reflection, from right-circular to left-circular (or vice versa), and that, therefore, the incident and reflected photons carry the same angular momentum $\hbar\hat{z}$ – which does not depend on the photon frequency $\omega_o$. Since no AM is transferred to the mirror, its rotational kinetic energy $\tfrac{1}{2}I_o\Omega^2$ remains the same before and after reflection. Consequently, the reflected photon will not suffer a Doppler shift, retaining the original frequency $\omega_o$ of the incident photon.

Alternatively, it is conceivable that, in consequence of the rotation of the mirror, the reflected photon either departs from pure circular polarization or acquires a certain amount of vorticity, so that its AM will deviate from $\hbar$. Let the angular momentum thus transferred to the mirror be $\alpha\hbar$, where the constant parameter $\alpha$ could be either positive or negative. The rotational kinetic energy of the mirror now changes to $\tfrac{1}{2}I_o(\Omega+\alpha\hbar/I_o)^2$, namely, a change of $\alpha\hbar\Omega$ from the initial value of $\tfrac{1}{2}I_o\Omega^2$, provided, of course, that $I_o$ is sufficiently large for the second-order term to be neglected. The reflected photon must thus be Doppler-shifted by $\Delta\omega=\alpha\Omega$ in order to account for the change in the rotational kinetic energy of the mirror. A simple theoretical way to determine the value of $\alpha$ does not appear to exist.

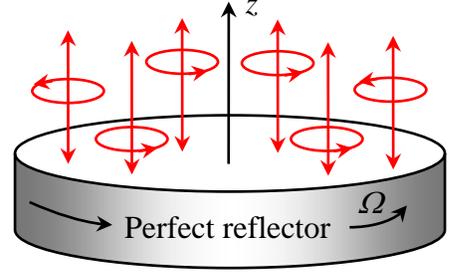

**Fig. 2**. A perfect flat reflector, having mass $M_o$ and moment of inertia $I_o$ around the $z$-axis, rotates around $z$ with the constant angular velocity $\Omega$. The mirror is in free space, and is illuminated at normal incidence by a circularly-polarized light pulse containing photons of energy $\hbar\omega_o$ and spin angular momentum $\hbar$.

Supposing that the angular velocity $\Omega$ of the mirror is fairly small and that, therefore, a given point on the mirror surface located a distance $R$ from the $z$-axis may be assumed to move at a constant linear velocity $V=R\Omega$, let us examine the reflection process using the methods of Special Relativity. Consider a monochromatic plane-wave of frequency $\omega_o$, propagating in free space in the $xyz$ frame shown in Fig. 3. The direction of propagation is specified by the $k$-vector $\boldsymbol{k}=(\omega_o/c)(\sin\theta\hat{\boldsymbol{x}}-\cos\theta\hat{\boldsymbol{z}})$. This beam is reflected from a flat mirror that is located in the $xy$-plane and moves with constant velocity $V$ along the $x$-axis. In the $xyz$ frame, the incident beam's phase-factor is given by

$$\exp[\mathrm{i}(\boldsymbol{k}\cdot\boldsymbol{r}-\omega_o t)]=\exp[\mathrm{i}(\omega_o/c)(x\sin\theta-z\cos\theta-ct)]. \tag{4}$$

Using the Lorentz transformation of space-time coordinates, we find the incident phase-factor in the $x'y'z'$ frame to be

$$\exp\{\mathrm{i}(\omega_o/c)[\gamma(x'+Vt')\sin\theta-z'\cos\theta-c\gamma(t'+Vx'/c^2)]\}$$

$$=\exp\left\{\mathrm{i}\gamma[1-(V/c)\sin\theta](\omega_o/c)\left[\frac{(\sin\theta-V/c)x'-(\gamma^{-1}\cos\theta)z'}{1-(V/c)\sin\theta}-ct'\right]\right\}. \tag{5}$$

Here, as usual, the Fitzgerald-Lorentz contraction factor is defined as $\gamma=1/\sqrt{1-(V/c)^2}$. Upon reflection from the stationary mirror, the beam's phase-factor in the $x'y'z'$ frame will become

$$\exp\left\{\mathrm{i}\gamma[1-(V/c)\sin\theta](\omega_o/c)\left[\frac{(\sin\theta-V/c)x'+(\gamma^{-1}\cos\theta)z'}{1-(V/c)\sin\theta}-ct'\right]\right\}. \tag{6}$$

Returning now to the *xyz* frame with a second Lorentz transformation, we will have, for the phase-factor of the reflected plane-wave,

$$\exp\left\{i\gamma[1-(V/c)\sin\theta](\omega_o/c)\left[\frac{(\sin\theta - V/c)\gamma(x-Vt)+(\gamma^{-1}\cos\theta)z}{1-(V/c)\sin\theta} - c\gamma(t-Vx/c^2)\right]\right\}$$
$$= \exp[i(\omega_o/c)(x\sin\theta + z\cos\theta - ct)]. \qquad (7)$$

The reflected beam thus suffers no Doppler shift and retains its orientation angle $\theta$, just as it would have, had the mirror depicted in Fig. 3 been stationary. It thus appears that the spinning mirror of Fig. 2 should *not* cause any Doppler shift of the reflected beam, as the individual segments of that mirror may be imagined to be moving at constant, albeit different, linear velocities. Bear in mind that the above argument is probably valid only at small angular velocities. Accurate experimental tests or a solution of Maxwell's equations in a rotating coordinate system will be needed to determine the validity of our conclusions at large values of the angular velocity $\Omega$.

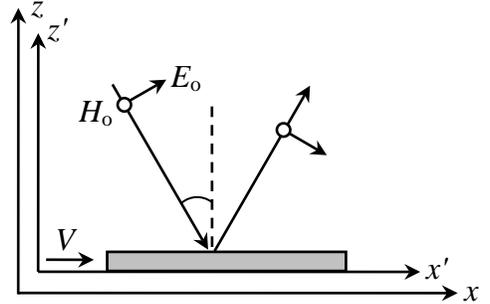

**Fig. 3**. A flat mirror moving along the *x*-axis at the constant velocity *V* is illuminated by a monochromatic plane-wave of frequency $\omega_o$. The plane of incidence is *xz*, and, in the *xyz* frame, the incident *k*-vector makes an angle $\theta$ with the *z*-axis. The rest frame of the mirror is denoted by *x'y'z'*.

**4. Absorption of circularly-polarized light within a spinning absorber**. Let a circularly-polarized photon of frequency $\omega_o$ enter a weakly-absorbing dielectric cylinder from free space, and proceed to get fully absorbed along the length of the cylinder, as shown in Fig. 4. The solid cylinder of mass $M_o$ and moment of inertia $I_o$ rotates around the *z*-axis with the constant angular velocity $\Omega$. Upon entering the absorber, the photon is Doppler-shifted to a new frequency $\omega = \omega_o \pm \Omega$, with the $\pm$ sign depending on whether the sense of circular polarization is the same as or opposite to that of $\Omega$. An intuitive understanding of this Doppler shift may be achieved if the *E*-field vector of each incident ray, ordinarily rotating around the ray axis at the optical frequency $\omega_o$, is seen from the perspective of an observer who rotates with the absorptive cylinder. The apparent rotation rate of the *E*-field vector is going to be $\omega_o \pm \Omega$, depending on whether the two rotations are in the same or in opposite directions. The frequency of the circularly-polarized beam as seen by the rotating observer is thus going to be the same as the angular velocity that the observer ascribes to the *E*-field vector.

Upon absorbing a photon, the slab accelerates (or decelerates) and, by the time the photon is fully absorbed, it reaches an angular velocity $\Omega + \Delta\Omega$. The change in the rotational kinetic energy of the cylinder is thus given by $\tfrac{1}{2}I_o(\Omega + \Delta\Omega)^2 - \tfrac{1}{2}I_o\Omega^2 \approx I_o\Omega\Delta\Omega$. Setting this equal to the change in the photon energy, $\pm\hbar\Omega$, then yields $I_o\Delta\Omega = \pm\hbar$. The photon AM of $\pm\hbar$ is thus seen to be transferred to the slab. The same argument will also work if the slab is initially stationary, provided that the Doppler shift $\Delta\omega$ is taken to be equal to one-half of its final angular velocity.

One may be tempted to apply the above argument even if the spinning cylinder does not absorb any light at all. The naïve argument goes as follows. Once the photon gets inside the slab, the AM of the combined system (cylindrical slab + photon) must change by $\pm\hbar$. This requires a change in the rotational kinetic energy of the combined system, which is then compensated by the Doppler shift of the photon. The problem with this argument is that it is not easy to separate the EM field from the material medium under such circumstances. Typically, a fraction of the EM angular momentum remains in the fields, while the rest is transferred to the medium as mechanical AM. So long as the photon remains inside, the cylinder rotates at a modified angular velocity due to its newly acquired mechanical momentum. Only when the photon leaves the cylinder at the end of its journey, will it be possible, in principle, to determine if and by how much the angular velocity of the cylinder has changed, and also what its average angular velocity has been during the time interval when the photon resided within the cylinder.

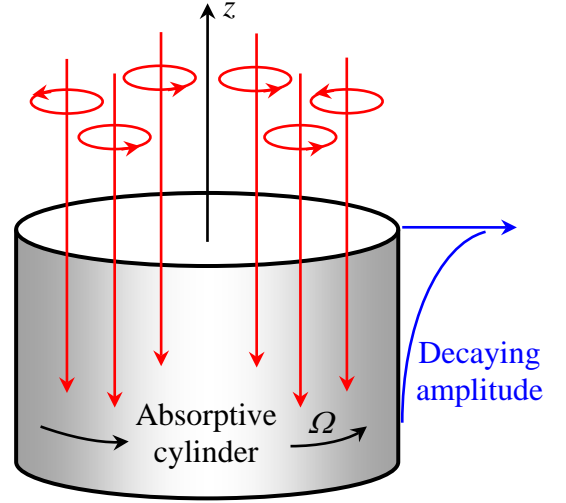

**Fig.4**. A weakly-absorptive solid dielectric cylinder is illuminated by a light pulse of frequency $\omega_o$, arriving from free space at normal incidence on the upper surface. The incident beam is circularly polarized, with each photon carrying energy $\hbar\omega_o$ and angular momentum $\pm\hbar$ along the $z$-axis. The cylinder, having mass $M_o$ and moment of inertia $I_o$, rotates around $z$ at a constant angular velocity $\Omega$. The absorber and the incidence medium are impedance matched to avoid reflection at the entrance facet. The cylinder height being greater than the penetration depth of the photons, the incident light pulse is absorbed in its entirety within the cylinder. It is assumed that $M_o$ is large enough to enable one to ignore the kinetic energy $\tfrac{1}{2}M_oV^2$ acquired by the cylinder in consequence of the absorption of photons with a linear momentum of $\hbar\omega_o/c$. The Doppler shift of the photons upon entering the cylinder is, therefore, entirely attributable to the change in the rotational kinetic energy $\tfrac{1}{2}I_o\Omega^2$ of the cylinder.

**5. Absorptive slab submerged in a transparent, homogeneous, isotropic dielectric medium**. Let an isotropic medium of complex refractive index $n_1+i\kappa_1$ be submerged in a transparent dielectric host of refractive index $n_o$, as shown in Fig.5. The Fresnel reflection coefficient at the interface is given by $\rho=(n_o-n_1-i\kappa_1)/(n_o+n_1+i\kappa_1)$. Inside the host dielectric, the $E$- and $H$-field amplitudes are $E_o\hat{y}$ and $H_o=(n_oE_o/Z_o)\hat{x}$, where $Z_o=\sqrt{\mu_o/\varepsilon_o}$ is the impedance of free space. Immediately inside the absorber, the field amplitudes are $E_1=(1+\rho)E_o\hat{y}$ and $H_1=(1-\rho)H_o\hat{x}$. The time-averaged Poynting vector is thus given by $<S>=\tfrac{1}{2}\mathrm{Re}(E_1\times H_1^*)=-(1-|\rho|^2)(\tfrac{1}{2}n_oE_o^2/Z_o)\hat{z}$. The fraction of the incident photons that enters the absorber is thus found to be $1-|\rho|^2=4n_on_1/[(n_o+n_1)^2+\kappa_1^2]$. Since each photon carries its spin angular momentum of $\pm\hbar$ into the absorber, the AM transferred to the absorber is $\pm 4n_on_1N\hbar/[(n_o+n_1)^2+\kappa_1^2]$, where $N$ is the number of incident photons. The ratio of the spin AM picked up by the submerged (isotropic) absorber to that picked up by the same absorber in the air (where $n_o=1$) is thus given by $n_o[(1+n_1)^2+\kappa_1^2]/[(n_o+n_1)^2+\kappa_1^2]$. For a typical set of parameters such as $n_o=1.5$, $n_1=2.0$, and $\kappa_1=0.1$, the submerged slab absorbs 98% of the incident photons, whereas the same slab in the air captures 89% of the photons. The difference in the AM picked up by the absorber in these two situations is thus only about 10%, which may not be easy to detect in an experimental setting. This is roughly what was observed in the experiments reported in [14], with the caveat that the experiment was done at microwave frequencies ($f=9.18$ GHz), with a dipole antenna acting as an absorber.

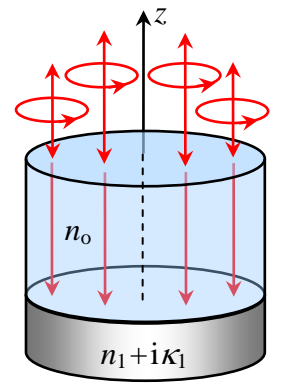

**Fig.5**. An absorptive slab of refractive index $n_1+i\kappa_1$ is submerged in a transparent dielectric of refractive index $n_o$. The circularly-polarized incident light pulse arrives at normal incidence at the entrance facet of the liquid, which is assumed to be impedance-matched to free space. Each absorbed photon will transfer an angular momentum of $\pm\hbar$ to the slab.

**6. Circularly-polarized light passing through a spinning transparent slab.** Shown in Fig. 6 is a circularly-polarized photon of frequency $\omega_o$ and angular momentum $\pm\hbar$ entering from free space into a spinning transparent slab of mass $M_o$, moment of inertia $I_o$, and angular velocity $\Omega$. The photon then emerges from the exit facet of the slab into free space. The slab is homogeneous and isotropic, with both its facets anti-reflection coated to prevent reflection losses when the light enters and exits. One may argue that, in passing through the slab, the photon's AM does not change and that, therefore, the slab neither gains nor loses any mechanical AM. Under such circumstances, the photon does not suffer any Doppler shift in consequence of its passage through the spinning slab.

Alternatively, one might argue that the emerging light pulse could have acquired some vorticity, or its state of circular polarization may have changed. Suppose the change in the photon AM is such that the slab's AM is required to have changed by $\alpha\hbar$, where $\alpha$ is a (positive or negative) constant. The change in the rotational kinetic energy of the slab, $\tfrac{1}{2}I_o(\Omega+\alpha\hbar/I_o)^2 - \tfrac{1}{2}I_o\Omega^2 \approx \hbar\alpha\Omega$, thus demands that the emergent photon be Doppler-shifted by $\Delta\omega = \alpha\Omega$. Needless to say, this type of argument cannot lead to a determination of the value of $\alpha$, which requires either experimental tests or a solution of Maxwell's equations in a rotating coordinate system.

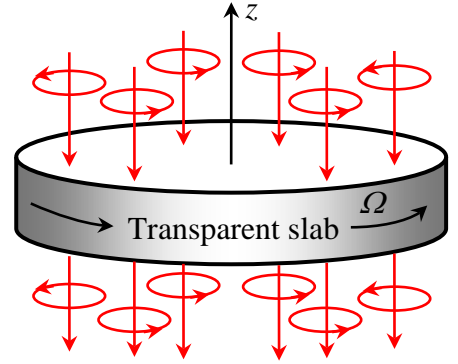

**Fig. 6**. A circularly-polarized light pulse of frequency $\omega_o$, containing photons of angular momentum $\pm\hbar$, passes through a spinning transparent slab in free space. The slab has mass $M_o$, moment of inertia $I_o$, and angular velocity $\Omega$. Anti-reflection coatings applied to the entrance and exit facets of the slab ensure the passage of the pulse without reflection losses. If the AM of the emergent pulse happens to be the same as that of the incident pulse, the angular velocity of the slab will remain the same before and after passage, and the photons will not suffer any Doppler shift in the process. If, however, the transmitted pulse acquires some vorticity, or if its state of polarization is modified, the slab's angular velocity will change, and there will be a corresponding Doppler shift in the frequency of the transmitted photons.

**7. Circularly-polarized light passing through a spinning half-wave plate.** A circularly-polarized photon of frequency $\omega_o$ enters a rotating transparent half-wave plate, as shown in Fig. 7. The plate has mass $M_o$, moment of inertia $I_o$, and angular velocity $\Omega$. The emergent photon should be Doppler-shifted to $\omega = \omega_o \pm 2\Omega$. This is because when the photon enters the slab it has one sense of circular polarization, but, immediately before exiting, it has the opposite sense. The initial Doppler shift of the photon, which occurs when the photon first enters the slab, is now doubled upon exiting (rather than being cancelled out, as in the case of an isotropic slab discussed in the preceding section).

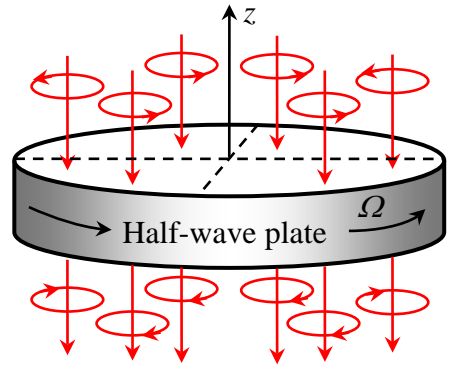

**Fig. 7**. A circularly-polarized light pulse of frequency $\omega_o$, containing photons with an AM of $\pm\hbar$, passes through a spinning half-wave plate in free space. The plate has mass $M_o$, moment of inertia $I_o$, and angular velocity $\Omega$; anti-reflection coatings applied to both its facets ensure the passage of the pulse without reflection losses. Upon passage through the slab, the polarization state of the photon goes from left- to right-circular (or vice versa). Consequently, $2\hbar$ of AM per photon is picked up by the slab, causing a change of $\pm 2\hbar\Omega$ per photon in its rotational kinetic energy. The corresponding Doppler shift $\Delta\omega$ of the transmitted photons is, therefore, $\pm 2\Omega$. Any deviation from $\Delta\omega = \pm 2\Omega$ would be an indication that either the transmitted pulse has acquired some vorticity, or that its polarization state is no longer purely circular.

The half-wave plate will accelerate (or decelerate) and, by the time the photon leaves the plate, its angular velocity reaches $\Omega + \Delta\Omega$. The change in the rotational kinetic energy of the slab, $I_o \Omega \Delta\Omega$, is now going to be equal to $\pm 2\hbar\Omega$, which means that the slab must have picked up (or lost) a total angular momentum $I_o \Delta\Omega$ equal to $2\hbar$. This is consistent with the fact that the photon's AM has gone from $+\hbar$ to $-\hbar$, or vice versa; see [5] for an early experimental affirmation of these ideas. Note that the emergent photon, by virtue of its Doppler shift, has now a different linear momentum as well. So, there must be a corresponding exchange of linear momentum, which requires an additional Doppler shift – unless $M_o$ is sufficiently large to render this secondary Doppler shift negligible.

**8. Concluding remarks**. While the interaction of light with material media moving at constant linear velocities can often be investigated with the help of Special Relativity, rotational motion and its effects on EM fields are much more complicated to analyze and harder to fathom. The Doppler shift phenomenon in conjunction with the conservation laws of energy, momentum, and angular momentum provides a simple analytical tool for the investigation of momentum exchange (linear as well as angular) between EM fields and material media. In our previous publications [11,12], we have shown the power of the Doppler-shift analysis in situations involving the exchange of linear momentum between EM fields and a material body moving at a constant linear velocity. In the case of rotating objects discussed in the present paper, however, the answers are not as clear-cut, primarily because the magnitude of the Doppler shift is not independently known in advance. To the extent that one can guess or estimate the Doppler shift resulting from the rotation of an object, one can speculate about the angular momentum exchanged in the interaction between the object and the EM fields. The analyses of the preceding sections have pointed out the intimate connection between the Doppler shift and the extent of AM exchange, so that the experimental determination of one could lead to a precise prediction of the magnitude of the other.